\documentclass{ajour}

\begin{document}


\title{Alternative approaches to the Casalbuoni-Brink-Schwarz Superparticle\footnote{
Submitted to Annals of Physics}}

\titlerunninghead{Approaches to the CBS Superparticle}
\authorrunninghead{Morten Nielsen, N. K. Nielsen}

\author{Morten Nielsen}
\email{morten@fysik.sdu.dk}
\author{N. K. Nielsen}
\email{nkn@fysik.sdu.dk}

\affil{Fysisk Institut, Syddansk Universitet, Odense Universitet, Denmark.}

\date{\today}


\begin{abstract}

 Wigner's method of induced representations is
applied to the $N=1$ super-Poincar\' e group, and by using a state
corresponding to the basic vector of the little group as a
Clifford vacuum we show that the spin operator of a supersymmetric
point particle obeys Wigner's constraints. As dynamical variables
for the particle we use canonical coordinates on the symmetry
group manifold. The physical phase space is then constructed using
a vielbein formalism. We find that the Casalbuoni-Brink-Schwarz
superparticle appears as a special case of our general
construction. Finally, the theory is reformulated as a gauge
theory where the gauge freedom corresponds to the choice of spin
constraints or, equivalently, the free choice of relativistic
center of mass. In a special case the gauge symmetry reduces to
the well known $\kappa$-symmetry.

{\em PACS numbers: 12.60.Jv, 02.20.Qs}
\end{abstract}

\vskip2pc


\section{Introduction}

The Casalbuoni-Brink-Schwarz (CBS) \cite{Casalbuoni, Brink,
Siegel} superparticle has played an important conceptual role for
superstring theory as a source of new inspiration for the  solution of the
problems encountered in the attempt to quantize supersymmetric string theories in a
manifestly covariant manner. An extensive literature on the
quantization of the superparticle exists (see \cite{Lyakhovich}
and references given there). In these papers the quantization
including the construction of the physical Hilbert space is
carried out by means of the covariant Becchi-Rouet-Stora-Tyutin or
Batalin-Fradkin-Vilkovisky methods.

As we shall see in the present paper an alternative method for construction of the
superparticle state space is provided by
Wigner's method of induced representations
\cite{Wigner} applied to the $N=1$ super-Poincar\'{e} group. Our other main
concern  is the connection between the conventional formulation of the superparticle
and the basic understanding of spinning point particles provided by the
classical papers of  Pryce
\cite {Pryce} who showed that there is a freedom in the choice of spin
operator corresponding to the arbitrariness of the relativistic
center of mass.

For a spinning particle the total angular momentum is the sum of the
orbital angular
momentum $L^{\mu\nu}$ and the spin $S^{\mu\nu}$ which must obey some
constraints. In particular, for a massive particle
the Pryce constraints
\begin{equation}
S^{\mu\nu}P_\nu=0,\label{pc}
\end{equation}
with the nice feature of covariance, are equivalent to the Wigner constraints
\begin{equation}
(P^0+m)S^{0j}-P^lS^{lj}=0, \label{wc}
\end{equation}
which arise naturally by the method of induced representations of the
Poincar\' e group \cite{Wigner}.

Here we shall mainly deal with the connection between (\ref{pc})
and (\ref{wc}). The spin operators and the corresponding position operators are related by
linear transformations the details of which can be found in Appendix
\ref{spcon}. Remarkably, in the customary formulation of the superparticle
the constraints
(\ref{pc}) are realized automatically, and as a consequence the components of the
position
operator do not
commute, cf. Eq. (\ref{xwithx}). However, it was observed by Brink and Schwarz
\cite{Brink} that a redefinition is possible in such a
way that the components of the new position operator commute mutually.
Their transformation formula is reminiscent of Eq. (\ref{XfraX}) relating the position operators
in the two cases where (\ref{pc}) and (\ref{wc}) apply.

In an earlier publication \cite{Ulrich} it was shown that the freedom of
choice of spin constraints
actually can be viewed as a gauge symmetry. This gauge symmetry is
in the present paper extended to the supersymmetric point particle
in a generalized  version of the CBS particle that, however, is
 physically equivalent to the original one.
 This gauge symmetry reduces in a special case to the
$\kappa $-symmetry of Siegel \cite {Siegel}.

In \cite{Ulrich} it was also found that the dynamical
degrees of freedom of a non-supersymmetric point particle can be described
by canonical coordinates on the Poincar\'e group manifold.
A similar analysis of the supersymmetric
point particle is made below. The particle will in this case move on the
super-Poincar\' e group manifold  projected onto the physical
superspace by means of supervielbeins constructed in accordance with
the supermanifold formalism of De Witt
\cite{DeWitt}. In addition to the usual mass shell constraint and spin
constraints one must now impose a set of fermionic constraints on the supercoordinate.
We find that this system is equivalent to the  superparticle considered in
\cite{Casalbuoni, Brink}. Using a modified method of gauge unfixing
\cite{Vytheeswaran} it
is seen that the gauge freedom in the choice of the spin constraints found
for the non-supersymmetric particle is still present. Furthermore, it is demonstrated
that the gauge unfixed theory can be gauge fixed again in such a way
that the  Dirac brackets lead to the same commutation relations as the
method of induced representations.  Finally
it is indeed found that the transition from Eq. (\ref{pc}) to (\ref{wc})
as well as the redefinition of position variables of
Brink and Schwarz  (see \cite{Brink} Eq. (21)) constitute gauge transformations.

The analysis will be carried out in several steps.

First, in Sec. \ref{fact} a factorization of a general
super-Poincar\' e transformation into an ordinary Poincar\'{e}
transformation and a supertranslation is found. Next, in Sec. \ref{wigner},
the Wigner construction  of the representations of
the Poincar\' e group \cite{Wigner} is extended to the
super-Poincar\' e group by means of this factorization. The
Clifford vacuum method of Salam  and Strathdee \cite{Salam} is
used in the basic frame of the little group and we restrict
ourselves for simplicity to the case of a spinless Clifford
vacuum. In this way one finds that the constraints
(\ref{wc}) still occurs as a natural candidate for a set of spin
constraints. An explicit expression for the spatial part of the
spin operator is derived from the structure relations in Sec.
\ref{spinop}, and by use of the constraints (\ref{wc}) the rest of the
components of the spin operator is determined.

The superparticle is then in Sec. \ref{maspart} identified with a
particle moving on the super-Poincar\'{e} manifold. The
constraints on the spin operator are obtained by imposing a set of
constraints on the fermionic degrees of freedom. These fermionic
constraints give rise to a second class constraint algebra (see
\cite{Hanson} and references given there) and thus do not define a
gauge theory. In Sec. \ref{gaugetheory} we use
gauge unfixing \cite{Vytheeswaran}, where half of these
constraints are singled out to form a first class algebra, and next gauge
fix the resulting gauge theory suitably, thus obtaining
a spin operator which obeys either one of the sets of constraints
(\ref{pc})-(\ref{wc}) or some other constraints depending on the
choice of gauge. We present two varieties of gauge unfixing the theory,
using projection operators constructed either by means of $\gamma ^5$ (in four dimensions) or
the free massless Dirac operator, and in each case construct Dirac brackets.
The appendices contain details on vielbeins and
on commutators involving the spin operator subject to the
constraints (\ref{pc}) and (\ref{wc}), respectively.

We use a metric $\eta _{\mu \nu }={\rm diag }(-1, 1, \cdots ,1)$
and the Dirac matrices are in a Majorana representation, with
\begin{equation}
\{\gamma _\mu ,\gamma _\nu \}=-2\eta _{\mu \nu }. \label{Dirr}
\end{equation}
 The charge
conjugation matrix is $-\gamma^0$ and the number of space-time
dimensions is denoted $D$.

\section{Factorizations} \label{fact}

A general Poincar\'e transformation can be factorized into a
translation and a Lorentz transformation. Such a factorization is
necessary in order to use Wigner's method of induced
representations \cite{Wigner} on a semidirect product group. This
factorization can be obtained just by applying two succesive
Poincar\'e transformations on a general vector. It is, however,
instructive to see how the factorization can be obtained from a
construction involving left and right vielbeins, since the
superspace part of a super-Poincar\'{e} transformation in a
similar way can be factorized from the ordinary Poincar\'{e}
transformation. We use the general formulation of supermanifolds
developed in \cite {DeWitt}. Definitions of and explicit
expressions for the vielbeins can be found in the appendices
\ref{superlie} and \ref{vielbeins}.
\subsection{Factorization of the Poincar\'{e}
transformation} \label{poincarefactor}
The Poincar\'{e} group consists of translations and Lorentz
transformations. The generators of infinitesimal transformations
are $P_\mu$ and $M_{\mu\nu}$, respectively. The generators fulfil
the commutation relations
\begin{eqnarray}
  \left[M_{\mu\nu},M_{\lambda\kappa}\right] =
   \frac{i}{2} C^{\xi\eta}_{\;\;\;\;\mu\nu,
   \lambda\kappa}M_{\xi\eta },
  \hspace{2mm}
  \left[M_{\mu\nu},P_\lambda\right] = iC^\xi_{\;\;\mu\nu,\lambda}P_\xi ,
  \label{SUSY1} \hspace{2mm}
  [P_\mu,P_\nu]= 0 \label{SUSY4}
  \end{eqnarray}
where the structure constants are
\begin{eqnarray}
&&C^\xi\hspace{0.11 mm}_{\mu\nu,\sigma}=\delta^\xi_{\;\;\nu}\eta_{\mu\sigma} -
    \delta^\xi_{\;\;\mu}\eta_{\nu\sigma} \label{SUSY5}
\\&&
C^{\xi\eta}_{\;\;\;\;\mu\nu,\sigma\rho} = \delta^\xi_{\;\;\sigma}
    C^\eta_{\;\;\mu\nu,\rho}-\delta^\eta_{\;\;\sigma}C^\xi_{\;\;\mu\nu,\rho}-
    \delta^\xi_{\;\;\rho}C^\eta_{\;\;\mu\nu,\sigma}+
    \delta^\eta_{\;\;\rho}C^\xi_{\;\;\mu\nu,\sigma}
\label{SUSY6}
\end{eqnarray}

Poincar\'{e} group elements are specified by $(a,\lambda)$, their
canonical coordinates in the sense of \cite{DeWitt}. Here $a$ and
$\lambda$ correspond to translations and Lorentz transformations,
respectively. A Lorentz transformation $\Lambda$ is given by
$\lambda ^{\mu \nu }$ through
\begin{eqnarray}
\Lambda^\mu_{\;\;\nu} = \left( e^{-C\cdot\lambda} \right)^{\mu }_{\;\;\nu }
\label{lt}
\end{eqnarray}
with
\begin{equation}
(C\cdot\lambda)^\mu _{\;\;\nu}
=\frac{1}{2}C^\mu\hspace{0.1 mm}_{\rho \sigma,\nu}\lambda ^{\rho \sigma }.
\end{equation}

Let composition of Poincar\'e group elements be given by the function $F$
according to
\begin{eqnarray}
(a_1,\lambda_1)\cdot (a_2,\lambda_2) = F[(a_1,\lambda_1),(a_2,\lambda_2)].
\end{eqnarray}
If $(da,0)$ is an infinitesimal translation, we get by a Taylor expansion,
using (\ref{left}):
\begin{eqnarray}
F[(da,0),(0,\lambda)]=
  (0,\lambda)+da\cdot u^{-1}[\lambda ]
\end{eqnarray}
that by means of (\ref{uer}) explicitly is
\begin{eqnarray}
F[(da,0),(0,\lambda)]=
  (da^{\mu}u^{-1}[\lambda ]_\mu \hspace{0.1 mm}^\nu,\lambda ^{\rho \sigma }). \label{inf}
\end{eqnarray}
This formula is only valid when an infinitesimal translation is
considered. However, when the product of two infinitesimal
transformations is applied group associativity allows one to use
(\ref{inf}) twice in succesion:
\begin{eqnarray}
F[(2da,0),(0,\lambda)]= F[(da,0),F[(da,0),(0,\lambda)]]= (2da\cdot
u^{-1}[\lambda ],\lambda). \label{ga}
\end{eqnarray}
Repeating this procedure $n$ times gives us
\begin{eqnarray}
F[(nda,0),(0,\lambda)]=(nda\cdot u^{-1}[\lambda ],\lambda)
\end{eqnarray}
Taking  $n\rightarrow\infty$ with $nda$  fixed,
$nda = a\cdot u[\lambda ]$, this becomes
\begin{eqnarray}
(a,\lambda) =
  F[(X,0),(0,\lambda)] \label{byt}
\end{eqnarray}
where we define $X^\lambda=a^\nu u_\nu^{\;\;\lambda}$
as the physical translation vector.

We do the same thing for $F[(0,\lambda),(da,0)]$:
\begin{eqnarray}
F[(0,\lambda),(da,0)]=
  (0,\lambda)+da\cdot v^{-1}[\lambda ]
\end{eqnarray}
with $v_\mu \hspace{0.1 mm}^\nu=u_\mu
\hspace{0.1 mm}^\rho\Lambda ^\nu\hspace{1 mm}_\rho$ the right vielbein.
By the same procedure as was used above we get:
\begin{eqnarray}
(a,\lambda)
  = F[(0,\lambda),(\Lambda X,0)]. \label{byt2}
\end{eqnarray}
By equating (\ref{byt}) and (\ref{byt2}) one finds the well known
factorization.

\subsection{Factorization of the super-Poincar\'e
transformation} \label{superpoincarefactor}

The $N=1$ super-Poincar\'{e} group is an extension of the Poincar\' e
group. The algebra is enlarged by the generator of infinitesimal
supertranslations $Q$
which is a Majorana spinor.
The canonical coordinate corresponding to a supertranslation is a Grassman
variable denoted $\xi ^{\alpha }$. New structure relations are
\begin{eqnarray}
  \left[M_{\mu\nu},Q_\alpha\right] = -\frac{i}{4}([\gamma_\mu,\gamma_\nu])_\alpha^
  {\;\;\beta}Q_\beta ,
  \hspace{2mm}
  \left\{Q_\alpha,Q_\beta\right\} = -2(\gamma \cdot P\gamma ^0)_{\alpha\beta},
  \hspace{2mm}[Q_\alpha,P_\mu] = 0
\nonumber \\ \label{SUSY3}
  \end{eqnarray}

Following the procedure of sec. \ref{poincarefactor}
one can factorize a supersymmetry transformation into a Poincar\'{e}
transformation and a supertranslation.
If $(d\xi,0,0)$ is an infinitesimal supertranslation a Taylor
expansion leads to the following identity
\begin{eqnarray}
F[(d\xi,0,0),(\xi,a,\lambda)]=(\xi,a,\lambda)+d\xi\cdot u^{-1}[(\xi,a,\lambda)].
\end{eqnarray}
Comparing this to the explicit expressions for the vielbeins found
in appendix (\ref{vielbeins}) it is seen that the multiplication
of a general group element by an infinitesimal supertranslation
affects the translation and supertranslation parts, while the
Lorentz part is unaffected. Using the group associativity in a way
similar to (\ref{ga}) one finds
\begin{eqnarray}
  &&\hspace{-6mm} F[(2d\xi ,0,0),(0 ,a,\lambda)] =
  F[(d\xi ,0,0),F[(d\xi ,0,0),(0,a,\lambda)]]
  \nonumber \\ &&
  \hspace{-6mm}=
  \left(2d\xi^\beta u^{-1}[\lambda]_\beta^{\;\;\alpha},a^\mu+d\xi^\beta
  u^{-1}[(0,\lambda)]_\beta^{\;\;\mu}
  + d\xi^\beta u^{-1}[(d\xi\cdot u^{-1},\lambda)]_\beta^{\;\;\mu},
  \lambda\right).
\end{eqnarray}
Repeating this process $n$ times gives
\begin{eqnarray}
F[(nd\xi,0,0),(0,a,\lambda)]=
  \left(nd\xi^\beta u^{-1}[\lambda]_\beta^{\;\;\alpha},a^\mu
  +d\xi^\beta
  \sum_{k=0}^{n-1}u^{-1}[(kd\xi\cdot u^{-1},\lambda)],\lambda\right)\nonumber \\
\label{above}
\end{eqnarray}
Now let $n\rightarrow\infty$ and $d\xi\rightarrow 0$ with $nd\xi$
fixed: $nd\xi^\alpha =\xi^\beta u[\lambda]_\beta^{\;\;\alpha}$.
Then Eq. (\ref{above}) reads
\begin{eqnarray}
(\xi ,a ,\lambda)=
  F\left[\left(\xi^\alpha u[\lambda]_\alpha^{\;\;\beta},0,0\right),(0,a^\mu+\frac{1}{2}
  \xi^\alpha
  u[(\xi^\gamma,\lambda)]_\alpha^{\;\;\nu}u^{-1}[\lambda]_\nu^{\;\;\mu},\lambda)\right]
  \label{sueq1}
\end{eqnarray}
Introducing the spacetime
translation vector $X$ and the supertranslation spinor $\theta$ by
\begin{eqnarray}
X^\lambda = a^\nu u_\nu^{\;\;\lambda} + \frac{1}{2} \xi^\beta
u_\beta^{\;\;\lambda},\hspace{2mm}
\theta^\alpha = \xi^\beta u_\beta^{\;\;\alpha} \label{theta}
\end{eqnarray}
we find
\begin{eqnarray}
&&(\xi,a ,\lambda)=F[(\theta,0,0), (0,X^\nu(u^{-1})_\nu \hspace{0.1 mm}^\mu,\lambda)]
  \label{st1}
\end{eqnarray}
Thus, the super-Poincar\'{e} transformation has been factorized into a Poincar\'{e}
transformation followed by a supertranslation.

The same procedure can be applied to the right vielbeins:
\begin{eqnarray}
F[(\xi,a,\lambda),(d\xi,0,0)]=
  \left(\xi^\alpha+d\xi^\beta v^{-1}[\lambda]_\beta^{\;\;\alpha},a^\mu+d\xi^\beta
  v^{-1}[(\xi,\lambda)]_\beta^{\;\;\mu},\lambda\right)
\end{eqnarray}
with $v$ denoting right vielbeins, and by the same procedure as before we get
\begin{eqnarray}
(\xi,a,\lambda)=
  F\left[(0,a^\mu+\frac{1}{2} \xi^\alpha v[(\xi,\lambda])_\alpha^{\;\;\rho}v^{-1}
  [\lambda]_\rho^{\;\;\mu},\lambda),(\xi^\alpha v[\lambda]_
  \alpha^{\;\;\beta},0,0)\right].
\label{sueq}
\end{eqnarray}

Under a Lorentz transformation given by the canonical coordinate
$\lambda ^{\mu \nu }$ the spinor transformation matrix is
\begin{eqnarray}
S^\alpha _{\;\;\beta} = \left( e^{-\frac{1}{8}\lambda ^{\mu \nu }[\gamma _\mu,
\gamma _\nu]} \right)^\alpha_{\;\;\beta},
    \label{lt2}
\end{eqnarray}
It is seen from Eq. (\ref{uer}) in connection with the definition of the structure
constants of Eq. (\ref{kommer})
that the following relation holds
\begin{equation}
v_\beta \hspace{0.1 mm}^{\alpha }=u_\beta \hspace{0.1 mm}^{\gamma }
S^\alpha\hspace{0.1 mm}_{\gamma }.
\end{equation}
By means of the identity (\ref{factor}) and
in terms of the translation vector and the
supertranslation spinor given in Eq. (\ref{theta}) we can rewrite Eq.
(\ref{sueq})
\begin{eqnarray}
(\xi,a,\lambda)=
  F\left[(0,X^{\rho}u^{-1}
  [\lambda]_\rho^{\;\;\mu},\lambda), (S^\alpha _{\;\;\beta}\theta ^\beta,0,0)\right].
\label{st2}
\end{eqnarray}
The factorizations of the super-Poincar\'{e} transformation
(\ref{st1}) and (\ref{st2}) are central for the construction of induced
representations of the
super-Poincar\'{e} group and are used for this purpose  in
section \ref{supertrans}.

\section{Spin constraints } \label{wigner}
The factorization of general Poincar\'{e} and super-Poincar\'{e} group elements
of (\ref{byt})-(\ref{byt2}) as well as (\ref{st1}) and (\ref{st2}) is used for
the construction of induced representations in the present section.
First we review how the Wigner constraints
(\ref{wc}) appears naturally for the Poincar\' e group when a group
theoretical analysis is made, and then it is shown how this analysis
carries over to the super-Poincar\' e group

\subsection{The Poincar\'{e} group}

Let a state vector be denoted by $|p,s\rangle$, where $p$ refers to
momentum quantum
numbers and $s$ to spin quantum numbers and other internal degrees of freedom.
Let a translation $X$ be represented
by an operator $T[X]$, a Lorentz transformation $\Lambda$ by an operator
$T[\Lambda]$ and a Poincare transformation $(a,\lambda)$ by an operator
$T[(a,\lambda)]$ in this space, where $a$ and $\lambda$ are canonical
coordinates, and $X^\nu=a^\mu u_\mu^{\;\;\nu}$ and $\Lambda=e^{-C\cdot\lambda}$
the associated translation and
Lorentz transformation, respectively. The operators $T[X]$, $T[\Lambda ]$ and
$T[(a, \lambda)]$
exist as unitary operators on the Hilbert space as demonstrated by Wigner \cite{Wigner, WignerII}.
The canonical coordinates $(a, \lambda )$ refer to
passive transformations, i.e. the system is unchanged but the observer is transformed.
In contrast, the Lorentz
transformation $\Lambda $ is an active transformation, where the system is transformed.

The little group \cite{Wigner, Ryder} corresponding to a fixed vector $q$,
$G_q$, is
the subgroup of the Lorentz group which leaves $q$ invariant.
For each momentum vector $p$ one singles out one Lorentz transformation
$\Lambda _{qp}$ with corresponding canonical coordinate
$\lambda _{qp}$, which transforms $q$ into $p$, and uses this transformation
to define a general state:
\begin{eqnarray}
\Lambda _{qp}q = p,\hspace{2mm}
|p,s\rangle \equiv T[\Lambda _{qp}]|q,s\rangle.
\label{pdef}
\end{eqnarray}
To an arbitrary Lorentz transformation $\Lambda$ corresponds a Wigner
transformation
\begin{eqnarray}
\Lambda_q[\Lambda,p] \equiv \Lambda ^{-1}_{q(\Lambda p)}\Lambda
  \Lambda _{qp} \in G_q \label{lamq}
\end{eqnarray}
which belongs to the little group. The general formula for the
representations of the Poincar\'{e} group is:
\begin{eqnarray}
T[(a,\lambda)]|p,s\rangle =
  e^{-iX\cdot (\Lambda ^{-1}p)}\sum_{s'}T_{s's}[\lambda_q[\Lambda^{-1},p]]|\Lambda ^{-1}
p,s'\rangle
  \label{genpoinc}
\end{eqnarray}
Thus all that is needed to perform a Poincar\'{e} transformation
on a state vector
is the representations of the little group.

\subsubsection{Massive case}\label{wigcon}

For a particle with non-zero rest mass $m$ the rest system vector,
$q^\mu = (m,\vec{0})$, is used as the basis of the Wigner analysis.
In this frame the little group $G_q$ consists of all spatial rotations, $G_q =
SO(D-1)$.
Boosts are
\begin{eqnarray}
\Lambda_{qp}  = \hat{\Lambda}(p) P
\label{boost}
\end{eqnarray}
with the definitions
\begin{eqnarray}
\hat{\Lambda}(p)^{\mu\nu} = \frac{2\tilde{p}^\mu\tilde{p}^\nu}{\tilde{p}^2} -
  \eta^{\mu\nu},\hspace{2mm}
\tilde{p}^\mu = (p^0+m,\vec{p})
\label{tuttelu}
\end{eqnarray}
and
with
$P =$ diag$(1,-1,-1,\cdots)$ the parity operator.
In the case of an infinitesimal Lorentz transformation with antisymmetric
parameter $\delta \lambda ^{\mu\nu}$
the transformation (\ref{lamq})
is an infinitesimal rotation with rotation
parameter
\begin{eqnarray}
\delta \hat{\lambda}^{ij} = \delta \lambda ^{ij} - \frac{p^i\delta \lambda
^j_{\;\;0}
-\delta \lambda ^j_{\;\;0}p^j}
{p^0+m}
\label{WIG}
\end{eqnarray}
The general expression for the transformation matrix of an
infinitesimal Lorentz transformation applied to a rest state is
\begin{eqnarray}
T_{s's}(\lambda_q[\Lambda,p]) = \delta _{s's} - \frac{i}{2}\delta \lambda
^{\mu\nu}
  (S_{\mu\nu})_{s's}
\end{eqnarray}
while application of the general formula (\ref{genpoinc}) gives:
\begin{eqnarray}
T_{s's}(\lambda_q[\Lambda,p]) = \delta_{s's} - \frac{i}{2}\delta \hat{\lambda}
  ^{ij}(S_{ij})_{s's}
\end{eqnarray}
where only spatial components of the spin operator are present.
Equating the two expressions using (\ref{WIG}) we obtain
\begin{eqnarray}
S_{j0} = \frac{p^i}{p^0+m}S_{ij} \label{wigncon1}
\end{eqnarray}
which in operator form are the Wigner constraints (\ref{wc}).

\subsubsection{Massless case}

For a massless particle a similar analysis can be carried out using
$q^\mu=(E, E, 0, \cdots, 0)$ as the basis for the Wigner analysis. In this case
the little group consists of
$D-2$ dimensional rotations as well as some combinations
of rotations and boosts. The generators
of these boost/rotation components of the little group is denoted $K_i$.
Introducing light cone coordinates as
$p^\pm=\frac{1}{\sqrt{2}}(p^0\pm p^{1})$
and correspondingly for the spin operator
this operator must fulfil
\begin{eqnarray}
S_{-i}=-\frac{1}{\sqrt{2}} K_i,\hspace{2mm}
S_{+i}=\frac{1}{p^+} (p^jS_{ij}-p^-S_{-i}-\sqrt{2}ES_{-i}), \hspace{2mm}
S_{+-}=\frac{p^iS_{-i}}{p^+} \nonumber \\ \label{masslesswc}
\end{eqnarray}
Because of the boosts contained in the $K_i$ generators the little group of the
massless particle is not compact, so all unitary representations are
infinite-dimensional or trivial. Therefore, in order to have a
finite-dimensional unitary representation we must demand that the
generators $K_i$ vanishes.
In this case the spin constraints become
\begin{eqnarray}
S_{-i}=S_{+-}=0, \hspace{2mm}
S_{+i}=\frac{1}{p^+}p^jS_{ij}.
\label{m0spin}
\end{eqnarray}

\subsection{The super-Poincar\'{e} group} \label{supertrans}

An analysis similar to the one of the previous section can be carried out for the super-Poincar\'{e}
group (cp. also \cite{Cornwell}).

If (\ref{lt2}) is examined in the case of a pure boost giving (\ref{pdef})
we find that the corresponding spinor transformation is given by
\begin{eqnarray}
&&{\rm massive\hspace{2 mm} case:   }
\nonumber \\&&(S_p)^\alpha_{\;\;\beta} =
  \frac{1}{\sqrt{2m(p^0+m)}} \left( \delta^\alpha_{\;\;\beta}(p^0+m) +
  (\gamma^0\vec{\gamma}\cdot\vec{p})^\alpha_{\;\;\beta}\right)
  \label{massivespinorpart}\\&&
  {\rm massless \hspace{2 mm}case:   }
  \nonumber \\&&
  (S_p)^\alpha_{\;\;\beta} =
  \frac{1}{\sqrt{2E(p^0+p^{1})}} \left( \delta^\alpha_
  {\;\;\beta}(p^0+E) +E
  (\gamma^0\gamma^{1})^\alpha_{\;\;\beta}+
  (\gamma^0\vec{\gamma}\cdot\vec{p})^\alpha_{\;\;\beta}\right)
  \nonumber \\&&
  \label{masslessspinorpart}
\end{eqnarray}

In the rest frame of a massive particle or the light-cone frame of a massless
particle the general structure relations reduce considerably.
We then use the $Q_q$  operators to carry out a
supertranslation on a particle state according to \cite{Salam, Wess}:
\begin{eqnarray}
T[\theta]|q, s\rangle = \sum_{s'}\left(e^{-i \bar{Q}_q\theta}\right)_{s's}
|q, s'\rangle
=\sum_{s'}R_{s's}(\theta)|q,s'\rangle \label{insp2}
\end{eqnarray}
In this way one can construct supermultiplets from a Clifford vacuum
characterized by its innate spin. Using the results (\ref{st1}) and (\ref{st2})
one gets for a general state
\begin{eqnarray}
T[\theta]|p,s\rangle = T[\theta]T[\Lambda_p]|q,s\rangle =
  T[\Lambda_p]T[ S_p ^{-1}\theta]|q,s\rangle.
\end{eqnarray}
This is a supertranslation on a
particle in the rest frame,
followed by a Lorentz transformation, so the effect of a supertranslation
on a general state is given by
\begin{eqnarray}
T[\theta]|p,s\rangle =
  \sum_{s'}R_{s's}( S_p^{-1}\theta)|p,s'\rangle .
  \label{sut}
\end{eqnarray}
Here it is important to note three things:
\begin{enumerate}
\item The effect of the
supertranslation is to bring the particle into a linear
combination of the elements of the supermultiplet to which the particle
belongs.
\item The coefficients of this linear combination $R_{s's}$ are
defined in the rest frame of a massive particle or the light-cone frame of a
massless
particle. So even when these coefficients appear in the
supertranslation of a general state we have to go to one of these frames to
evaluate
the coefficients.
\item Eq. (\ref{insp2})
implies
\begin{equation}
R_{s's}(S^{-1}_p\theta ) = \left(e^{-i\bar{Q}_qS_p^{-1}\theta }\right)_{s's}=
\left(e^{-i\bar{Q}_p\theta }\right)_{s's}
\end{equation}
so when supertranslating a general state instead of a basic state one can change
$\theta$
as in (\ref{sut}) or one can use boosted $Q$s instead of the basic frame $Q_q$s.
\end{enumerate}

According to (\ref{st1})  the operator representing a
general super-Poincar\'e transformation factorizes into
\begin{eqnarray}
T[(\xi ,a ,\lambda)] = T[\theta]T[(0,X^\nu(u^{-1})_\nu \hspace{0.1 mm}^\mu,\lambda)]
\end{eqnarray}
and acts consequently on a general quantum mechanical state $|p,s\rangle $
as follows:
\begin{eqnarray}
T[(\xi,a,\lambda)]|p,s\rangle =
  e^{-iX\cdot\Lambda ^{-1} p}\sum_{s',s''}T_{s's}(\lambda[\Lambda ^{-1},p])
  R_{s''s'}(S_{\Lambda ^{-1}p}^{-1}\theta )|\Lambda ^{-1}p,s''\rangle
\nonumber \\
\label{gensup}
\end{eqnarray}
where (\ref{genpoinc}) has been used.

We can compare this result, accomplished by use
of the factorization (\ref{st1}), with the general
Poincar\'e transformation in Eq.
(\ref{genpoinc}). The new thing is the appearance of the $R_{s's}$
coefficients. However, we still have the $T_{s's}$ coefficients
appearing in exactly the same way leading again to the Wigner constraints
(\ref{wc}) for the spin operator, but this time the spin operator operates
within a supermultiplet. Especially, the supermultiplet may involve a Clifford
vacuum with spin 0, in which case the construction leads to the
Casalbuoni-Brink-Schwarz superparticle.

\section{The spin operator} \label{spinop}

Having verified in the previous section that the spin operator when constructed by
the method of induced representations obeys the
constraints (\ref{wc}) also in the supersymmetric case, we are now ready for an explicit construction of
the spin operator, using the structure relations of the little group basic frame.

\subsection{The massive particle}

To determine $S_{\mu\nu}$ one uses the nontrivial structure relations (\ref{SUSY3}) in the rest frame:
\begin{eqnarray}
&&[S_{ij},Q_{r}]=-\frac{i}{4}[\gamma_i,\gamma_i]
  Q_{r}, \label{MQcomre}
\\
&&\{Q_{r\alpha},Q_{r\beta}\} =
2m\delta _{\alpha \beta }, \label{QQcomre}
\end{eqnarray}
where $Q_r$ is the supersymmetry generator in the rest frame.
From Eqs. (\ref{MQcomre})-(\ref{QQcomre}) one determines the
following spin
operator:
\begin{eqnarray}
S^{ij}=-\frac{i}{8m}\bar{Q}_{r}\gamma^{0ij}Q_{r}
\end{eqnarray}
with $\gamma^{\mu \nu \lambda }=\gamma^{[\mu }\gamma^\nu
\gamma^{\lambda ]}$ the completely antisymmetric product. The spin
operator can be expressed in terms of the boosted $Q$s by
inverting (\ref{massivespinorpart}). The outcome is:
\begin{eqnarray}
S^{ij} =
  -\frac{i}{8m^2}\bar{Q}_p\gamma^{ij\mu}Q_pp_\mu +
  \frac{i}{8m^2(p^0+m)}\bar{Q}_p(p^i\gamma^{0j\mu} - p^j\gamma^{0i\mu})p_\mu Q_p
\end{eqnarray}
Since this spin operator obeys the Wigner constraints (\ref{wc})
the last components of the spin operator can also be determined.
Defining the new spin operator
\begin{eqnarray}
\hat{S}^{\mu \nu } = -\frac{i}{8m^2}\bar{Q}_p\gamma^{\mu \nu
\lambda }Q_pp_\lambda \label{boostspin}
\end{eqnarray}
which obviously fulfils the Pryce constraints (\ref{pc}) we can
incorporate also the time components of this spin operator such
that
\begin{equation}
S^{\mu \nu }= \hat{S}^{\mu \nu }-\frac{1}{p^0+m}p^\mu
\hat{S}^{0\nu}+
  \frac{1}{p^0+m}p^\nu \hat{S}^{0\mu}.
\label{transtre}
\end{equation}
This is the spin operator transformation formula
describing a transition between a system where the Wigner
constraints are valid and the same system where the Pryce
constraints are valid, cf. eqs. (\ref{transet})-(\ref{transto}).

To establish the connection of $S^{\mu \nu }$ to the spin operator of
\cite{Brink}
one introduces the coordinate $\theta $ according to
\begin{eqnarray}
Q_p
=  2i\gamma\cdot p\theta.
\end{eqnarray}
The anticommutator
$\{Q_{p\alpha },Q_{p\beta }\}=2(\gamma\cdot p\gamma^0)_{\alpha\beta}$
then leads to
\begin{eqnarray}\{\theta_{\alpha},\theta_{\beta}\}
=  \frac{1}{2m^2}(\gamma\cdot p \gamma ^0)_{\alpha\beta}
\label{kuku}
\end{eqnarray}
and the spin operator (\ref{boostspin}) becomes
\begin{eqnarray}
\hat{S}^{\mu\nu}=
-\frac{i}{2}\bar{\theta}\gamma^{\mu\nu\lambda}\theta P_\lambda .
\label{smyny}
\end{eqnarray}
These formulas can be compared to ref. \cite{Brink} and are recognized as
the anticommutation relations
of the superspace  coordinates and the expression giving the spin operator
in the case of a massive superparticle. Thus the massive version of the CBS superparticle
has been constructed by means of the method of induced representations since the
state space was determined in Sec. \ref{spinop}.

\subsection{The massless particle}

In the light-cone system of a massless particle where $p^\mu = (E, E,
0, \dots)$ the two nontrivial structure relations (\ref{SUSY3}) reduce to
\begin{eqnarray}
&&\{Q_{lc\alpha},Q_{lc\beta}\} =
2\sqrt{2}E(\gamma ^-\gamma^0)_{\alpha \beta }\label{QQcomre1}
, \\&&
[S_{ij},Q_{lc}]=-\frac{i}{4}[\gamma_i,\gamma_i] Q_{lc},
\label{MQcomre1}
 \\&&
[S_{-j},Q_{lc}] =
  -\frac{i}{2}\gamma_-\gamma_jQ_{lc}.
\end{eqnarray}
In these expressions $Q_{lc}$ is the supersymmetry generator in the light-cone
system and the indices  $i, j$ are in the range $2, \cdots, D-1$.
In order to express the spin operator in terms of $Q_{lc}$ one must
require
\begin{equation}
Q_{lc}=\frac{1}{\sqrt 2}\gamma^-\gamma^0 Q_{lc} \label{projprop}
\end{equation}
thus halving the number of independent components of $Q_{lc}$. A consequence is
\begin{equation}
S_{-i}=0.
\end{equation}
From Eqs. (\ref{QQcomre1})-(\ref{MQcomre1}) one finds:
\begin{eqnarray}
S^{ij}=\frac{i}{32E}Q_{lc}
  [\gamma^i,\gamma^j]Q_{lc}=\frac{i}{16\sqrt{2}E}\bar{Q}_{lc}
  \gamma^{ij+}Q_{lc}.
\end{eqnarray}
The remaining components of the spin operator are fixed by the constraints
(\ref{m0spin}).

To find the boosted $Q$'s one uses Eq.
(\ref{masslessspinorpart}) and Eq. (\ref{projprop}):
\begin{eqnarray}
Q_{p} =
  -\frac{1}{2\sqrt{Ep^+}}\gamma \cdot p   \gamma^0 Q_{lc}.
\label{booost}
\end{eqnarray}
By comparison with \cite {Brink}
one realizes that $\gamma ^0Q_{lc}$ is proportional
to the $S$-variable of that paper. The method thus also has allowed
construction of the massless superparticle.
\section{Particle on a group manifold} \label{maspart}
Having accomplished the construction of the CBS superparticle by means of the
method of induced representations, we next use a different starting point to
elucidate the relationships between the two sets of spin constraints (\ref{pc}) and (\ref{wc}).
In the course of the construction, the spin operator had to be transformed
according to Eq. (\ref{transtre}) (cf. Eq. (\ref{transto})). For
a non-supersymmetric particle with spin, it was demonstrated in \cite{Ulrich}
that this transformation can be considered a gauge transformation. However,
this identification is not possible within the framework developed so far.

Following \cite{Hanson, Ulrich} we identify the superparticle with a particle moving
on the super-Poincar\'{e} group manifold.  First
the naive action for a free particle moving on the supergroup manifold is
considered. Next the connection between our construction and the
Casalbuoni-Brink-Schwarz superparticle is made.

\subsection{Naive action}\label{action}

In analogy to the case of the Poincar\'{e} group \cite{Hanson, Ulrich}
 we now imagine a supersymmetric particle with spin moving on
the super-Poincar\'{e} group manifold, where we use the general
formulation of supermanifolds developed in
\cite {DeWitt}. The group manifold is parametrized by the canonical coordinates
$a^\mu$, $\lambda^{\mu\nu}$ and $\xi^\alpha$. We project the corresponding canonical momenta
onto the physical generators according to (\ref{projektion}) with the vielbeins
listed in Appendix \ref{vielbeins}:
\begin{eqnarray}
&&\Pi_\mu = P_\nu u_\mu^{\;\;\nu}  , \hspace{4 mm} \Pi_\alpha
   = \left(\bar{Q}_\beta
  u_\epsilon^{\;\;\beta} + P_\nu u_\epsilon^{\;\;\nu}\right)
  \left(\gamma^0\right)^\epsilon_{\;\;\alpha}, \nonumber \\
  &&\Pi_{\mu\nu}=\frac{1}{2} M_{\rho\sigma}u_{\mu\nu}^{\;\;\;\;\rho\sigma} +
  P_\lambda u_{\mu\nu}^{\;\;\;\;\lambda} + \bar{Q}_\alpha
  u_{\mu\nu}^{\;\;\;\;\alpha}
   \label{supimp}
\end{eqnarray}
to ensure the correct structure relations,
while coordinates are
\begin{eqnarray}
X^\lambda = u_\nu^{\;\;\lambda}a^\nu + \frac{1}{2}
  u_\beta^{\;\;\lambda}\xi^\beta,
  \hspace{4 mm}\theta^\alpha = u_\beta^{\;\;\alpha}\xi^\beta .
  \label{supcoor}
\end{eqnarray}
The naive action on the group manifold is
\begin{eqnarray}
S_{\rm naive}=\int d\tau (\Pi_\mu\dot{a}^\mu + \frac{1}{2}
\Pi_{\mu\nu}\dot{\lambda}^{\mu\nu} +
  \bar{\Pi}\dot{\xi} ).
\label{megetnaiv}
\end{eqnarray}
Now one uses the Cartan-Maurer equations (\ref{Cartan-Maurer})
as well as (\ref{supimp}),
(\ref{supcoor}) and
\begin{eqnarray}
P_{\theta }\equiv
  Q-i\gamma\cdot P\theta.   \label{standard}
\end{eqnarray}
Here $P_{\theta }$ is the conjugate momentum to $\theta $, with
$\{\theta ^{\alpha },P_{\theta }^{\beta }\}_{PB}=\left(\gamma^0\right)
^{\alpha \beta }$ with the subscript $PB$ denoting Poisson brackets (Poisson brackets
involving Grassmann variables are defined in Eq. (\ref{grasspoin})).
After lengthy calculations one obtains the naive action
\begin{eqnarray}
S_{\rm naive} = \int d\tau\left( P_\mu \dot{X}^\mu+
   \bar{P}_{\theta }\dot{\theta}
   +\frac{1}{2} \Sigma_{\mu \nu }\sigma^{\mu \nu }
  \right)
\label{naiv}
\end{eqnarray}
where
\begin{equation}
\sigma ^{\mu \nu}=(\Lambda ^{-1})^{\mu
}\hspace{0.1 mm}_{\lambda }\dot{\Lambda }^{\lambda \nu
}=\frac{1}{2}\dot{\lambda }^{\rho \sigma
}u_{\rho \sigma }\hspace{0.1 mm}^{\mu \nu }
\end{equation}
and
where the spin of the Clifford vacuum appears
\begin{eqnarray}
\Sigma_{\mu \nu }  =
  M_{\mu \nu } - X_\mu P_\nu + X_\nu P_\mu
  - S_{\mu\nu} \label{sigma}
\end{eqnarray}
as we would expect.  Here
\begin{equation}
S_{\mu \nu}=-\frac{1}{4}\bar{\theta }
[\gamma ^{\mu }, \gamma ^{\nu }]P_{\theta}
\label{spin}
\end{equation}
is the part of the
angular momentum which comes from the Grassmann part of superspace. We still
have to impose constraints on $S_{\mu \nu}$ in order to reduce the number
of independent components .

\subsection{Connection to the CBS superparticle}

In order to regain the Casalbuoni-Brink-Schwarz superparticle  we change the coordinates such that
the spin operator fulfils the Pryce constraints and assume a spinless Clifford
vacuum ($\Sigma _{\mu \nu }=0$).
Comparing (\ref{standard}) to  \cite{Casalbuoni, Brink} we require
\begin{eqnarray}
\psi=P_{\theta}-i\gamma\cdot P\theta=0 \label{bscon}
\end{eqnarray}
so the generator becomes
\begin{eqnarray}
Q=P_{\theta }+i\gamma\cdot P\theta=
  2i\gamma\cdot P\theta.
\end{eqnarray}
The anticommutator $\{Q_\alpha,Q_\beta\}=2(\gamma\cdot P\gamma^0)_{\alpha\beta}$
then leads to
\begin{eqnarray}
\{\theta_{\alpha },\theta_{\beta}\}=
  -\frac{1}{2P^2}(\gamma\cdot P \gamma ^0)_{\alpha \beta }
\label{thetatheta}
\end{eqnarray}
and the spin operator (\ref{boostspin}) becomes
\begin{eqnarray}
\hat{S}^{\mu\nu}=
  -\frac{i}{2}\bar{\theta}\gamma^{\mu\nu\lambda}\theta P_\lambda.
\label{smunu}
\end{eqnarray}
The action is obtained from (\ref{naiv}) with $\Sigma=0$ and with constraint
terms added:
\begin{eqnarray}
S= \int d\tau\left( P_\mu \dot{X}^\mu+
   \bar{P}_{\theta }
  \dot{\theta}-\bar{\lambda}(P_{\theta}-i\gamma\cdot P\theta)-e(P^2+m^2)\right)
\end{eqnarray}
with $m$ a mass parameter and $e$ and $\lambda $ Lagange multipliers. This
is recognized as the action of
the CBS superparticle in first order form, while
eqs. (\ref{thetatheta}) and (\ref{smunu}) are identical to eqs.
(\ref{kuku}) and (\ref{smyny}).

\section{Gauge theory of the superparticle} \label{gaugetheory}
The results obtained so far can be summarized in the following way:
 Using the method of
induced representations, we obtained in Eq. (\ref{smyny}) the spin
operator of the massive CBS superparticle, with the anticommutation
relations (\ref{kuku}) in the same form as in \cite{Casalbuoni, Brink}.
In the massless case the CBS superparticle also emerged from the method of
induced representations. Next it was shown how the superparticle also could
be interpreted as moving on the super-Poincar\'{e} group manifold, when
the set of constraints
(\ref{bscon}) is imposed upon the naive action of a free particle.

In this connection it is puzzling that the superparticle spin operator
obeys Eq. (\ref{pc}). When determining the spin operator
by the method of induced representations we had to carry out a
redefinition according to Eq.
(\ref{transtre}) to obtain this relation, while it was ensured by the set of constraints (\ref{bscon})
when the superparticle was considered moving on the super Poincar\'{e}
manifold.

In the case of a nonsupersymmetric point particle it is known that
the arbitrariness in the constraints of the spin operator as
reflected in the mutually exclusive conditions given in Eqs.
(\ref{pc})-(\ref{wc}) reflects a deep symmetry of a particle with
spin (or an extended object) related to the arbitrariness of the
relativistic center of mass \cite{Pryce}. Furthermore, it has
been demonstrated that this symmetry can be formulated as a gauge
symmetry such that Eq. (\ref{transtre}) expresses a gauge
transformation \cite{Ulrich}. This raises the interesting
possibility that an equivalent formulation of the CBS
superparticle exists, such that it has the same physical contents
but allows a gauge symmetry corresponding to the transformation
(\ref{transtre}). In this relation it should be mentioned that a
gauge symmetry of the massless CBS superparticle was found by Siegel \cite{Siegel}
 and has given rise to an extensive literature on the covariant
quantization problem for this theory (see \cite{Lyakhovich} and
references quoted there). What we shall determine below
is a more general scheme containing Siegel's gauge symmetry
as a special case.

The constraints
(\ref{bscon}) are second class while the constraints of a gauge
theory are first class. The task at hand consists of halving the number
of these constraints  in such a way that those remaining  are first class
and thus define a
 gauge theory. This gauge theory should reduce to the CBS superparticle
 in a particular gauge.  The Dirac brackets of the super-Poincar\'{e} generators
should be unaffected by the choice of gauge.
   The procedure of obtaining a gauge theory from a theory defined by second class
constraints  is known as gauge unfixing \cite{Vytheeswaran}.

\subsection{General framework}

Gauge unfixing on the set of constraints of Eq.
(\ref{bscon}) is carried out by means of two projection operators
$Y_\pm$, in such a way that one obtains the new constraints:
\begin{equation}
\psi _{\pm }=Y_\pm
(P_\theta-i
\gamma \cdot P_{\pm }\theta )=0
\label{Ycon}
\end{equation}
where $P_+\neq P_- $, and only the constraints $\psi _+$ are kept, while
$\psi _-$ are taken as gauge fixing conditions. The constraints $\psi _+$ should
\begin{enumerate}
\item be first class:
\begin{equation}
\{\psi _+, \psi _+\}_{PB}=0,
\end{equation}
\item have weakly
vanishing Poisson brackets with the generators of the super Poincar\'{e} group
in order
to ensure the correct algebra of the generators after Dirac quantization:
\begin{equation}
\{M_{\mu \nu }, \psi_+\}_{PB} \simeq 0,\hspace{2mm}\{P_{\mu },
\psi_+\}_{PB} \simeq 0,\hspace{2mm}
\{Q , \psi _+ \}_{PB}\simeq 0.
\label{gaugeinv}
\end{equation}
\end{enumerate}
In order to obtain Eq. (\ref{gaugeinv}) one has to choose $P_+=P$.
On the other hand $P_-$ is arbitrary and a particular choice means
fixing the gauge freedom. If one chooses the gauge $P_-=P$ the
resulting model is thus identical to the CBS superparticle.

Having obtained a first class constraint algebra one can determine
gauge transformations of a general variable $A$.
 The generator of an infinitesimal gauge transformation with gauge parameter
$\lambda $ is
\begin{equation}
Q=\bar{\lambda }\psi _+,
\label{gaugeet}
\end{equation}
and $A$ transforms according to
\begin{equation}
\delta A=\{Q, A\}_{PB}
\label{gaugeto}.
\end{equation}
Eq. (\ref{gaugeinv}) then implies that the generators $M_{\mu \nu
}$, $P_\mu $ and $Q$ are gauge invariant.

Two sets of projection operators $Y_\pm$ are considered.
Projection using chiral constraints is the simplest one in terms
of the algebra involved, but the succes relies on an antisymmetric
$\gamma^{D+1}$. Hence this procedure is only useful in some
dimensionalities e.g. four dimensions, but not ten dimensions. The
other set of projection operators involves the free massless Dirac
operator and works in any number of dimensions. However, these
projection operators may be ill defined at $P^2\rightarrow 0$ and
hence cannot immediately be applied to massless particles.

On the mass shell the spin constraints should be given by
\cite{Ulrich}
\begin{equation}
S^{\mu \nu }(P+P_-)_{\nu}=0 \label{gencon}
\end{equation}
which e.g. gives each of the sets of constraints (\ref{pc})-(\ref{wc}) with
the proper choice of $P_-$. The aim of this section is to show
that there exists a choice of gauge where (\ref{gencon}) reduce to
the Wigner constraints (\ref{wc}) and where we obtain Dirac
brackets (see \cite{Hanson} and references given there) involving
position and spin operators consistent with the commutation
relations obtained directly when the spin operator is assumed to
obey the Wigner constraints (\ref{wc}) as per the induced
representation theory. These commutators are given in Eqs.
(\ref{hurraet})-(\ref{hurrato}).

This brings us to the main point of our construction. The
corresponding commutators obtained in \cite{Casalbuoni, Brink}
and given in Eqs. (\ref{xwithx})-(\ref{xwiths}) on the mass shell are
those that apply when the spin operator obeys the Pryce constraints
(\ref{pc}). Since gauges giving respectively the Pryce and the
Wigner constraints have been determined it is concluded in analogy with
the nonsupersymmetric case that the two versions of the theory
differ only by a gauge transformation.

\subsection{Chiral projections in four dimensions}

Consider first the case where the projection operators $Y_\pm$ are the chiral
projections such that
\begin{equation}
Y _{\pm }=\frac{1\pm \gamma ^5}{2}.
 \label{chi}
\end{equation}
In four dimensions each chirality separately has vanishing
Poisson brackets:
\begin{equation}
\{\psi _+, \psi _+\}=\{\psi _-, \psi _-\}=0.
\end{equation}
For $P_-$ we choose
\begin{equation}
P_-=\bar{P}=(m, \vec{0}).
\label{pssi}
\end{equation}
\subsubsection{Dirac quantization}

Dirac brackets of arbitrary variables $A$ and $B$ are
\begin{eqnarray}
\{A, B\}'=\{A, B\}_{PB}-\{A, (\psi _+)_\alpha \}_{PB}
(C^{-1}_{+-})^{\alpha \beta }\{(\psi _-) _\beta ,B\}_{PB}
\nonumber \\
-\{A, (\psi _-)_\alpha \}_{PB}
(C^{-1}_{-+})^{\alpha \beta }\{(\psi _+) _\beta ,B\}_{PB}
\end{eqnarray}
with
\begin{equation}
(C^{-1})_{-+} =((C^{-1})_{+-})^T =
-\frac{i}{(P+\bar{P})^2}Y_+\gamma ^0
\gamma \cdot(P+\bar{P}).
\end{equation}

The spin operator $S^{\mu \nu }$ as given
in (\ref{spin}) is according to the constraints (\ref{Ycon})
with the projection operators $Y_\pm$ specified in (\ref{chi}):
\begin{equation}
S^{\mu \nu }\simeq S_V^{\mu \nu }-
A^\mu (P-\bar{P})^{\nu }+A^\nu (P-\bar{P})^{\mu },
\end{equation}
where
\begin{equation}
S_{V}^{\mu \nu }\simeq-\frac{i}{4}\bar{\theta }
[\gamma ^{\mu },
\gamma^{\nu }]\gamma \cdot \frac{P+\bar{P}}{2}
\theta,\hspace{3mm}
A^\mu =\frac{i}{4}\bar{\theta }\gamma ^\mu \gamma ^5\theta.
\end{equation}
In the present case
we get, using that $\theta $ is a Majorana spinor:
\begin{equation}
S^{\mu \nu }(P+\bar{P})_{\nu}=-A^{\mu }(P^2-\bar{P}^2)+A\cdot
(P+\bar{P})(P-\bar{P})^{\mu}
\end{equation}
that does not vanish even for $P^2=\bar{P}^2$.
Thus, to obtain constraints of the form (\ref{gencon}) one has to modify
$S^{\mu \nu }$. This modification amounts to
\begin{equation}
\hat{S}^{\mu \nu }=S^{\mu \nu }+2(\bar{P}^{\mu }P^{\nu }-\bar{P}^{\nu }P^{\mu })
\frac{A\cdot (P+\bar{P})}{(P+\bar{P})^2}
\label{modig}
\end{equation}
that only affects $S^{\mu \nu}$ for $\mu =0$ or $\nu =0$, and
for which
\begin{equation}
\hat{S}^{\mu \nu }(P+\bar{P})_{\nu}=-\left(A^\mu-(P+\bar {P} )^\mu\frac{A\cdot
(P+\bar{P})}{(P+\bar{P})^2}\right ) (P^2-\bar{P}^2).
\end{equation}
Dirac brackets for the spin operator (\ref{modig}) are
\begin{equation}
\{\hat{S}^{\mu \nu }, \hat{S}^{\lambda \rho }\}'=\Delta ^{\mu \lambda}
    \hat{S}^{\nu \rho }
  -\Delta ^{\nu \lambda }\hat{S}^{\mu \rho }+\Delta ^{\nu \rho }\hat{S}^{\mu
    \lambda }
  -\Delta ^{\mu \rho }\hat{S}^{\nu \lambda }
\label{SoSi}
\end{equation}
with
\begin{equation}
\Delta ^{\mu \nu }=\eta^{\mu \nu }-2\frac{P^{\mu }\bar{P}^{\nu }+P^{\nu }\bar{P}^{\mu }}
{(P+\bar{P})^2}. \label{dhat}
\end{equation}
Here was used
\begin{equation}
\{A^\mu , A\cdot(P+\bar{P}) \}'=\{S_{V}^{\mu \nu }, A\cdot (P+\bar{P})\}=0.
\end{equation}

Having redefined the spin operator $S^{\mu \nu }$ according to
(\ref{modig}) we have to redefine
the position operator $X^{\mu }$ also since the total Lorentz
transformation generator
$M_{\mu \nu }$ should be unmodified.
In this way $\hat{X}^{\mu }$ is fixed (apart from a term proportional to
$P^{\mu }$) to
\begin{equation}
\hat{X}^{\mu }=X^{\mu }-2\bar{P}^{\mu}\frac{A\cdot (P+\bar{P})}{(P+\bar{P})^2}.
\label{XNUL}
\end{equation}
The redefinition (\ref{XNUL}) only affects the time component of the position operator.
For the new position operator the following Dirac brackets are obtained
\begin{equation}
\{\hat{X}^{\mu }, \hat{X}^{\nu}\}'=4(\bar{P}^{\mu }P^{\nu}-\bar{P}^{\nu
}P^{\mu})
\frac{A\cdot (P+\bar{P})}{((P+\bar{P})^2)^2},
\end{equation}
\begin{equation}
\{\hat{X}^{\mu }, \hat{S}^{\nu \lambda }\}'=
2\frac{\bar{P}^{\nu }\hat{S}^{\lambda \mu }-
\bar{P}^{\lambda }\hat{S}^{\nu \mu }}{(P+\bar{P})^2}-\frac{4P^{\mu }}
{(P+\bar{P})^2}(\bar{P}^{\lambda }D^{\nu \rho }A_{\rho }
-\bar{P}^{\nu }D^{\lambda \rho }A_{\rho }).
  \nonumber \\
\end{equation}
For the spinorial coordinate we get
\begin{eqnarray}
\{\theta ^{\alpha }, \theta ^{\beta }\}^{\prime }
=-\frac{i}{(P+\bar{P})^2}(
\gamma \cdot(P+\bar{P})\gamma ^0 )^{\alpha \beta }.
\label{thetatehta}
\end{eqnarray}

\subsubsection{The mass shell constraint}

The final Dirac brackets are obtained by the mass shell constraint
\begin{equation}
\phi _1=P^2+m^2 \label{ms}
\end{equation}
and the corresponding gauge condition
\begin{equation}
\phi _2=\hat{X}^0-\tau \label{tid}
\end{equation}
in terms of which the final Dirac brackets of the variables $A$ and $B$ are
\begin{eqnarray}
\{A ,B \}^*=\{A ,B \}'-\{A ,\phi _1 \}'\frac{1}{2P^0}
\{\phi _2 ,B \}'+\{A ,\phi _2 \}'\frac{1}{2P^0}
\{\phi _1 ,B \}'.
\nonumber \\\label{stjerne}
\end{eqnarray}
In this way one obtains the Dirac brackets equivalent to the commutators
(\ref{hurraet}) and (\ref{hurrato})
obtained by
the method of induced representations:
\begin{equation}
\{\hat{X}^\mu ,P^\nu \}^*=\eta ^{\mu \nu }-\frac{P^{\mu }}{P^0}\eta ^{0\nu },
\end{equation}
\begin{equation}
\{\hat{X}^{\mu }, \hat{X}^{\nu }\}^*=0,
\end{equation}
\begin{equation}
\{\hat{X}^\mu , \hat{S}^{\nu \lambda }\}^*=\{\hat{X}^\mu , \hat{S}^{\nu \lambda }\}'
-\frac{P^{\mu }}{P^0}\{\hat{X}^0 , \hat{S}^{\nu \lambda }\}',
\end{equation}
whence
\begin{equation}
\{\hat{X}^i , \hat{S}^{jk }\}^*=0,
\end{equation}
\begin{equation}
\{\hat{X}^i , \hat{S}^{0j
}\}^*=\frac{1}{P^0+m}\hat{S}^{ij}-\frac{P^i}{P^0(P^0+m)}\hat{S}^{0j},
\end{equation}
and finally
\begin{equation}
\{\hat{S}^{\mu \nu }, \hat{S}^{\lambda \rho }\}^*=
\{\hat{S}^{\mu \nu }, \hat{S}^{\lambda \rho }\}'
\end{equation}
in agreement with
Eqs. (\ref{hurraet})-(\ref{hurrato}).

\subsection{Projection by the Dirac operator} \label{Dirac}

Instead of the chiral projections in (\ref{chi}) consider the projection operators
\begin{equation}
Y_\pm=\frac12\left(1\pm\frac{i\gamma\cdot P}{\sqrt{P^2}}\right),
\hspace{1 mm}
\bar{Y}_\pm=\frac12\left(1\pm\frac{i\gamma\cdot \bar{P}}{\sqrt{P^2}}\right)
\label{Ypm}
\end{equation}
where $Y_\pm$ is used to define the new constraints and $\bar{Y}_\pm$ are
used to simplify the following calculations.
We  specify $P_-$ according to:
\begin{equation}
P_-=\bar{P}=(-i\sqrt{P^2},\vec{0}).
\label{psi}
\end{equation}
Eq. (\ref{psi}) in contrast to Eq. (\ref{pssi}) shows explicit dependence on the momentum
operator $P$. This is necessary in order to obtain the Dirac brackets algebra that
after quantization
leads to the commutation relations (\ref{hurraet}) and (\ref{hurrato}).

From Eqs.
(\ref{spin}) and (\ref{Ycon}) one finds in this case
\begin{equation}
S^{\mu\nu}
  \simeq \hat{S}^{\mu\nu}+\frac{1}{4\sqrt{P^2}}(P^\mu \bar{P}^\nu-P^\nu \bar{P}^\mu )
  \bar{\theta}\theta
\end{equation}
with
\begin{eqnarray}
&&\hat{S}^{\mu\nu}=-\frac{i}{4}\bar{\theta}\gamma^{\mu\nu\lambda}\theta
(P+\bar{P})_\lambda
-\frac{1}{4\sqrt{P^2}}\bar{\theta}\gamma^{\mu\nu\lambda\tau}
  \theta P_\lambda \bar{P}_{\tau} .
\end{eqnarray}

\subsubsection{Dirac quantization}

After finding and inverting the constraint algebra one finds
preliminary Dirac brackets for arbitrary variables $A$ and $B$:
\begin{eqnarray}
\{A,B\}'=\{A,B\}_{PB}+\frac{2iP^2}{(P\cdot(P+\bar{P}))^2}
  \{A,\psi _+\}_{PB}\gamma^0Y_-
  \gamma\cdot \bar{P}Y_+\{\psi_+,B\}_{PB}\hspace{3mm}
\nonumber \\
-\frac{\sqrt{P^2}}{P\cdot(P+\bar{P})}
  \left(\{A,\psi _+\}_{PB}\gamma^0Y_-\{\psi _-,B\}_{PB}
  -\{A, \psi _-\}_{PB}\gamma^0Y_+\{\psi _+,B\}_{PB}\right).
  \label{deq}
\nonumber \\
\end{eqnarray}
Dirac brackets for the spin operator are:
\begin{eqnarray}
&&\{S^{\mu\nu},S^{\rho\sigma}\}'-\{S^{\mu\nu},S^{\rho\sigma}\}
  =-
  \frac{P^\mu \bar{P}^\rho + \bar{P}^\mu P^\rho}{P\cdot(P+\bar{P})}
   \hat S^{\nu\sigma}
\nonumber \\ &&
  -\eta^{\mu\rho}\frac{1}{4\sqrt{P^2}}(P^\nu \bar{P}^\sigma-P^\sigma \bar{P}^\nu )
  \bar{\theta}\theta
    +\mbox{ permutations}.
\end{eqnarray}
This result indicates that one should consider instead the modified spin
operator $\hat S^{\mu\nu}$ for which:
\begin{eqnarray}
\{\hat S^{\mu\nu},\hat S^{\rho\sigma}\}'=\{S^{\mu\nu},S^{\rho\sigma}\}'
  =\Delta^{\mu\rho}\hat S^{\nu\sigma}
  -\Delta^{\nu\rho}\hat S^{\mu\sigma}+\Delta^{\nu\sigma}\hat S^{\mu\rho}-
   \Delta^{\mu\sigma}\hat S^{\nu\rho}
\nonumber \\ \label{SSS}
   \end{eqnarray}
where
\begin{equation}
\Delta^{\mu\rho}=\eta^{\mu\rho}
  -\frac{P^\mu \bar{P}^\rho + \bar{P}^\mu P^\rho}{P\cdot(P+\bar{P})}
\end{equation}
which is identical to (\ref{dhat}) on the mass shell.
The redefinition of the spin operator must be accompanied by a redefinition of
the position operator:
\begin{equation}
\hat X^\mu = X^{\mu }
  -\frac1{4\sqrt{P^2}}\bar\theta\theta\bar P^\mu .
\end{equation}
The  Dirac brackets involving the new position variable are
\begin{eqnarray}
\{\hat {X}^\lambda, \hat{S}^{ \rho\sigma }\}'=
  \frac{1 }{P\cdot(P+\bar{P})}
  \left(\eta^{\lambda\mu}+\frac{P^{\lambda }\bar{P}^\mu }{P^2}\right)
(\hat{S}_\mu \hspace{1 mm}^\rho\bar{P}^\sigma
-\hat{S}_\mu \hspace{1 mm}^\sigma\bar{P}^\rho).
\end{eqnarray}
\begin{equation}
\{\hat X^\mu,\hat X^\nu\}'
  =-\frac{1}{P\cdot(P+\bar P)}\frac{1}{4\sqrt{P^2}}(P^\mu \bar{P}^\nu-P^\nu \bar{P}^\mu )
  \bar{\theta}\theta .
\end{equation}
Finally, we find
\begin{eqnarray}
\{\theta^\alpha,\theta^\beta\}'=-\frac{P^2}{(P\cdot(P+\bar P))^2}
  \left(\left(i\gamma \cdot (P+\bar{P})
+\frac{[\gamma \cdot P, \gamma \cdot \bar{P}]}{2\sqrt{P^2}}
\right)\gamma^0\right)^{\alpha\beta}
\end{eqnarray}
that is different from Eq. (\ref{thetatehta}).

\subsubsection{The mass shell constraint}

The final Dirac brackets are found by the mass shell constraint
along with a  gauge fixing
condition according to Eqs. (\ref{ms}), (\ref{tid}) and (\ref{stjerne}).
For the position operator we find
\begin{equation}
\{\hat X^\mu, \hat X^\nu\}^*=0
\end{equation}
and for the spin operator the brackets are unchanged and given by Eq. (\ref{SSS}).
For the spatial components of the spin operator
\begin{equation}
\{\hat X^\lambda , \hat S^{ij}\}^*=0
\end{equation}
while for the remaining components of the spin operator we find
\begin{equation}
\{\hat X^\lambda,\hat S^{0j}\}^*=\frac{1}{P^0+m}
  \left(\hat S^{\lambda j}-\frac{P^\lambda}{P^0}\hat S^{0 j}\right).\label{xxx}
\end{equation}
It is again seen that the Dirac brackets (\ref{SSS}) and (\ref{xxx}) are
in accordance with the commutators (\ref{hurraet})-(\ref{hurrato}).

\subsubsection{Gauge Transformations}

With Eq. (\ref{gaugeto}) defining gauge transformations,
where $Y_\pm$ are fixed according to Eq. (\ref{Ypm}),
the resulting gauge symmetry is a generalization of the local fermionic
symmetry of the massless CBS superparticle discovered by
Siegel \cite{Siegel}.

The following gauge transformations are found
by insertion into (\ref{gaugeet}) and (\ref{gaugeto}):
\begin{equation}
\delta \theta =\bar{\lambda}Y_+\theta,
\end{equation}
\begin{eqnarray}
\delta \hat{X}^\mu=-\bar{\lambda}Y_+i\gamma ^\mu\theta+\bar{\lambda}Y_+\bar{Y}_-
\frac{1}{2}i(\gamma ^\mu-P^\mu \frac{\gamma \cdot P}{P^2})\theta-\frac{1}{2}P^\mu \frac{P\cdot\bar{P}}
{P^2\sqrt{P^2}}\bar{\lambda}Y_+\theta .
\nonumber \\
\label{xtrans}
\end{eqnarray}
Introducing here $\kappa =\frac{2}{\sqrt{P^2}}\lambda $, choosing the Pryce gauge
condition $\bar{P}^\mu =P^\mu $ and finally using the mass shell condition $P^2=0$
one regains the gauge symmetry of \cite{Siegel}. Here it should be noted
that the position operator can always be redefined
by addition of a term proportional to the momentum operator. This
amounts to a gauge transformation
generated by the mass shell constraint.

Another consequence of Eq. (\ref{xtrans}) follows in the special case
\begin{equation}
\lambda =\theta \delta t
\end{equation}
with $\delta t$ an infinitesimal real parameter. In this case Eq. (\ref{xtrans}) reduces to
\begin{eqnarray}
\delta \hat{X}^\mu=\frac{i}{8P^2}\delta t\bar{\theta}
\gamma ^{\mu \nu \lambda }\theta P_\nu \bar{P}_\lambda
\label{bspw}
\end{eqnarray}
where terms proportional to $P^\mu $ are disregarded. Eq. (\ref{bspw})
holds for any choice of $\bar{P}$ and
is the
infinitesimal version of Eq. (\ref{XfraX}) (with $\bar{P}$ specified in (\ref{psi})) and
of the Brink and Schwarz coordinate transformation formula (see \cite {Brink} Eq. (21))
for $\bar{P}=P^-$. The finite version of  Eq. (\ref{bspw}) can be obtained through
integration of infinitesimal gauge transformations.

Needless to say, similar considerations on gauge transformations can be made
in the case where chiral constraints are used for gauge unfixing.

\section{Conclusion and outlook}

Using the factorization of a general super-Poincar\' e
transformation, which was carried out by means of the vielbein
formalism, we have applied the method of induced representations
to the $N=1$ super-Poincar\' e group. By combining this
with the Clifford vacuum method of Salam and Strathdee we have then
shown that the Wigner constraints for the spin operator occur in a
natural way. This allows one to find an explicit expression for
the spin operator using only the structure relations of the
super-Poincar\' e group, and the relation to the
Casalbuoni-Brink-Schwarz superparticle is demonstrated.

Next a different analysis was performed. The superparticle was
considered moving on the $N=1$ super-Poincar\' e group manifold.
By imposing the proper constraints the CBS particle is the result
of this. By the use of projection operators half of the
constraints could be selected to serve as the generators of gauge
transformations, while the other half was considered fixing the gauge.
It is immediately obvious how one should fix the gauge to recover
the CBS superparticle where the spin operator obeys the Pryce
constraints. Using Dirac quantization
we then showed that for another gauge choice the resulting
commutation relations corresponds to those expected if the Wigner
constraints are valid. By analogy to similar calculations for the
nonsupersymmetric case it is concluded that this is in fact a
gauge theory where the gauge freedom corresponds to the choice of
spin constraints or, equivalently, the free choice of relativistic
center of mass. We also showed how in a special case the gauge symmetry
reduces to the well known $\kappa$-symmetry.

One can imagine several interesting ways to generalize this work:
A Clifford vacuum with nonzero spin, $N=2$ supersymmetry, and
strings and branes.

{\bf Acknowledgment:} One of us (NKN) is grateful for an interesting
conversation on the topic of this paper with Professor P.
van Nieuwenhuizen some years ago.

\appendix{}
\section{Super Lie groups} \label{superlie}
Let $F^a[x,y]$ be the multiplication functional of a general supergroup
\cite{DeWitt}. Left- and right derivatives of $F$ are
\begin{eqnarray}
L^{\;\;a}_b[\bar{z}] \equiv  \left.
  \frac{\stackrel{\rightarrow}{\delta}}{\delta\bar{x}^b} F^a[\bar{x},\bar{z}]
  \right|_{\bar{x}=1},
\label{left}
\hspace{2 mm}
R^a_{\;\;b}[\bar{z}] \equiv \left.
   F^a[\bar{z},\bar{y}] \frac{\stackrel{\leftarrow}{\delta}}{\delta\bar{y}^b}
   \right|_{\bar{y}=1}
\end{eqnarray}
with the explicit representations
\begin{eqnarray}
\left ({L^T}^{-1}\right )^b_{\;\;a} =
  \left(\frac{e^{C\cdot\bar{x}}-1}{ C\cdot\bar{x}}\right )^b_{\;\;a}
=\int_0^1dt\left(e^{tC\cdot\bar{x}}\right)^b_{\;\;a}, \label{leftinv}\\
\left(R^{-1}\right )^b_{\;\;a} =
  \left(\frac{1-e^{-C\cdot\bar{x}}}
  {C\cdot\bar{x}}\right )^b_{\;\;a}
= \int_0^1dt\left(e^{-tC\cdot\bar{x}}\right )^b_{\;\;a} \label{rightinv}
\end{eqnarray}
where \footnote{In the sign factor
$(-1)^a$ corresponding to the quantity $A$  one ascribes to $a$ the value
$0$ for $A$ an ordinary
numbera and 1 for $A$ a Grassmann number.}
\begin{eqnarray}
(C\cdot x)^a_{\;\;b} = (-1)^{bc}x^cC^a_{\;\;cb}
\label{sign}
\end{eqnarray}
Here $C^a_{\;\;cb}$ are the supergroup structure constants and $x^c$ are
the canonical coordinates on the group manifold.
The supertranspose is defined by
\begin{equation}
\left(M^a_{\;\;b}\right)^T = (-1)^{b(a+b)}M_b^{\;\;a}. \label{supertransss}
\end{equation}
We introduce the
notation $u_a^{\;\;b}=(L^{-1})_a^{\;\;b}$ and $v_a^{\;\;b}=({R^T}^{-1})_a^{\;\;b}$.
The right vielbeins $v$  are obtained from the left vielbeins $u$ by the replacement
$x^a \rightarrow -x^a$.

Poisson brackets are in the presence of Grassmann variables defined according to
\begin{eqnarray}
\{A, B\}_{PB}=\sum _{\alpha ,\beta } ((A\frac{\stackrel{\leftarrow}{\partial}}
{\partial q^\alpha }
)\Gamma ^\alpha \hspace{0.1  mm}_\beta (\frac{\stackrel{\rightarrow}{\partial}}
{\partial p_\beta }B)
-(-1)^{A\cdot B}
(B\frac{\stackrel{\leftarrow}{\partial}}{\partial q^\alpha })
\Gamma ^\alpha \hspace{0.1  mm}_\beta
(\frac{\stackrel{\rightarrow}{\partial}}{\partial p_\beta }A)).
\nonumber \\
\label{grasspoin}
\end{eqnarray}
A derivative with respect to a coordinate is a right-derivative (denoted
$\frac{\stackrel{\leftarrow}{\partial}}{\partial q^{\alpha }}$)
while a derivative with respect to a momentum is a left-derivative (denoted
$\frac{\stackrel{\rightarrow}{\partial}}{\partial p_\beta }$). For Grassmann
variables this distinction is important.  The quantity $\Gamma ^\alpha
\hspace{0.1  mm}_\beta $ is equal to
$\delta^\alpha \hspace{0.1  mm}_\beta $ if $\alpha ,\beta $ refer to
ordinary numbers while one must require
$$
(\Gamma ^\alpha \hspace{0.1  mm}_\beta )^*=-\Gamma ^\alpha \hspace{0.1
mm}_\beta
$$
if $\alpha ,\beta $ refer to Grassmann variables.
$\Gamma ^\alpha \hspace{0.1  mm}_\beta $ should also be nonsingular but is
otherwise unrestricted.

With the  definition of Poisson brackets  given above the following
projection formula leads to group generators $I_\alpha $:
\begin{equation}
I_{\alpha }=\Pi _{\beta }(\Gamma ^{-1})^{\beta }\hspace{0.1 mm}_{\epsilon}
(u^{-1})_\delta\hspace{0.1 mm}^\epsilon K_{\alpha }\hspace{0.1 mm}^{\delta }
\label{projektion}
\end{equation}
with $K$ a nonsingular matrix that ensures that $I_\alpha $ is real if $\Pi
_\alpha $ is real.
The Cartan-Maurer equation is
\begin{eqnarray}
((u^{-1})_\alpha \hspace{0.1 mm}^\gamma )_{,\delta }(u^{-1})_{\beta }
\hspace{0.1 mm}^\delta
-(-1)^{\alpha \beta }
((u^{-1})_\beta \hspace{0.1 mm}^\gamma)_{,\delta}(u^{-1})_{\alpha
}\hspace{0.1 mm}^\delta
=(u^{-1})_\delta \hspace{0.1 mm}^\gamma C^{\delta }\hspace{0.1 mm}_{\alpha
\beta },
\nonumber \\
\end{eqnarray}
where the derivatives are right derivatives, or equivalently
\begin{equation}
(u_{\alpha }\hspace{0.1 mm}^{\gamma }),_{\beta }
-(u_{\beta }\hspace{0.1 mm}^{\gamma }),_{\alpha }
=-(-1)^{\epsilon (\beta +\delta )}C^{\gamma }\hspace{0.1 mm}_{\delta \epsilon }
u_{\alpha }\hspace{0.1 mm}^{\delta }u_{\beta }\hspace{0.1 mm}^{\epsilon }.
\label{Cartan-Maurer}
\end{equation}
The Cartan-Maurer equation leads to the structure relation of generators
\begin{equation}
\{I_{\alpha }, I_{\beta }\}=I_{\gamma }
(K^{-1})_{\epsilon }\hspace{0.1 mm}^{\gamma }
C^{\epsilon }\hspace{0.1 mm}_{\zeta \iota }
K_{\alpha }\hspace{0.1 mm}^{\zeta }
K_{\beta }\hspace{0.1 mm}^{\iota }
\label{III}
\end{equation}
where the structure constants $C^{\gamma}\hspace{0.1 mm}_{\alpha \beta}$
have the property
\begin{equation}
C^{\gamma}\hspace{0.1 mm}_{\alpha \beta}^*=(-1)^{\alpha \cdot \beta }
C^{\gamma}\hspace{0.1 mm}_{\alpha \beta}.
\end{equation}

\section{The Super-Poincar\'{e} group} \label{vielbeins}

For the $N=1$ super-Poincar\'e group we take $\Gamma =K=\gamma ^0$, with
$\gamma^0$ a
Dirac matrix from the Majorana representation constructed according to Eq.
(\ref{Dirr}).
By comparison of Eq. (\ref{III}) with the structure relations (\ref{SUSY3}) one
obtains the following structure constants complementing those of Eqs.
(\ref{SUSY5})-(\ref{SUSY6}):
\begin{equation}
C^{\mu }\hspace{0.1
mm}_{\alpha \beta }=-2i(\gamma ^0\gamma ^{\mu
})_{\alpha \beta },
\hspace{1 mm}
C^{\beta
}\hspace{0.1 mm}_{\mu \nu ,\alpha }
=-(\frac{1}{4}\gamma ^0[\gamma _{\mu }, \gamma _{\nu }]\gamma ^0)_{\alpha }
\hspace{0.1 mm}^{\beta }.
\label{kommer}
\end{equation}

The nontrivial left vielbeins are:
\begin{eqnarray}
u_\mu ^{\;\;\lambda} =
\int_0^1dt\left(e^{tC\cdot\lambda}\right)^\lambda_{\;\;\mu},
\hspace{2 mm}
u_{\mu\nu} ^{\;\;\;\;\lambda\kappa}  =
  \int_0^1dt\left(e^{tC\cdot\lambda}\right)^{\lambda\kappa}_{\;\;\;\;\mu\nu},
\hspace{2 mm}
 u_\alpha^{\;\;\beta} = \int_0^1dt\left(e^{tC\cdot\lambda}\right)^\beta_{\;\;\alpha}
 \nonumber \\
\label{uer}
\end{eqnarray}
as well as
\begin{eqnarray}
u_{\beta}^{\;\;\;\;\lambda} = \xi^\epsilon C^\lambda_{\;\;\alpha\delta}
  \int_0^1dt\int_0^tdu\left(e^{uC\cdot\lambda}\right)^\alpha_{\;\;\epsilon}
  \left(e^{tC\cdot\lambda}\right)^{\delta}_{\;\;\beta},
\end{eqnarray}
\begin{eqnarray}
u_{\mu\nu}^{\;\;\;\;\beta} = \frac{1}{2} \xi^\delta
  C^\beta_{\;\;\alpha ,\kappa\tau}\int_0^1dt\int_0^tdu
  \left(e^{uC\cdot\lambda}\right)^\alpha_{\;\;\delta}\left(
  e^{tC\cdot\lambda}\right)^{\kappa\tau}_{\;\;\;\;\mu\nu}
\end{eqnarray}
and
\begin{eqnarray}
u_{\mu\nu}^{\;\;\;\;\lambda} &=&
   \frac{1}{2} a^\xi C^\lambda_{\;\;\sigma ,\kappa\tau} \int_0^1dt\int_0^tdu
  \left(e^{uC\cdot\lambda}\right)^\sigma_{\;\;\xi}
  \left(e^{tC\cdot\lambda}\right)^{\kappa\tau}_{\;\;\;\;\mu\nu}
\\ & &
  +\frac{1}{2}\int_0^1dtt^2\int_0^1d\alpha_1\int_0^1 d\alpha_2\int_0^1d\alpha_3
  \delta(1-\alpha_1-\alpha_2-\alpha_3)\nonumber \\ & &
  \left(e^{t\alpha_1C\cdot\lambda}
  \right)^\lambda_{\;\;\sigma}(C\cdot\xi)^\sigma_{\;\;\alpha}
  \left(e^{t\alpha_2C\cdot\lambda}\right)^\alpha_{\;\;\gamma}(C\cdot\xi)
  ^\gamma_{\;\;\kappa\tau}\left(e^{t\alpha_3C\cdot\lambda}\right)^{\kappa\tau}
  _{\;\;\;\;\mu\nu}. \nonumber \label{nyu}
\end{eqnarray}
In the last expression one should remember the sign in the term
$(C\cdot\xi)^\sigma_{\;\;\alpha}$ specified according to (\ref{sign}).

From these expressions the corresponding right vielbeins $v$ are
obtained by a change of sign of the canonical coordinates. The
following identity applies:
\begin{equation}
\frac{1}{2}\xi^\alpha
u_\alpha^{\;\;\rho}{u^{-1}}_\rho^{\;\;\nu}=\frac{1}{2}\xi^\alpha
v_\alpha^{\;\;\rho}{v^{-1}}_\rho^{\;\;\mu}.
\label{factor}
\end{equation}

The Poincar\' e group vielbeins are the first two of
Eq. (\ref{uer}) as well as that of Eq. (\ref{nyu}) where
one should take $\xi=0$.

\section{ Spin constraints } \label{spcon}

The space components of the spin operator $S_{ij}$ obey the usual $SO(D-1)$
algebra:
\begin{equation}
[S^{ij},S^{kl}]=i(\delta ^{ik}S^{jl }+\delta ^{jl }S^{ik }
  -\delta ^{il }S^{jk }-\delta ^{jk }S^{il }).
\end{equation}
while the position operator commutes with the spin components
\begin{equation}
[X^{i},S^{jk}]=0.
\end{equation}
The remaining components are determined by the Wigner constraints:
\begin{equation}
S^{0i}=\frac{1}{P^0+m}P^jS^{ji}
\end{equation}
and we keep the mass-shell condition:
\begin{equation}
P^{2}+m^2=0
\end{equation}
Then we compute:
\begin{equation}
[S^{0i},
S^{jk}]=i\frac{1}{P^0+m}(P^jS^{ik}-P^kS^{ij})+i(\delta ^{ik}S^{0j}-\delta
^{ij}S^{0k})
\end{equation}
\begin{equation}
[S^{0i},
S^{0j}]=\frac{i}{(P^0+m)^2}\vec{P}^2S^{ij}+\frac{i}{P^0+m}(P^jS^{0i}
-P^iS^{0j})
\end{equation}
and
\begin{equation}
[S^{0i}, X^{j}]=i\frac{1}{P^0+m}S^{ij}+i\frac{1}{P^0(P^0+m)}P^jS^{0i}.
\label{hurraet}
\end{equation}
The commutation relations of the spin operator are summarized:
\begin{equation}
[S^{\mu \nu },S^{\lambda
\rho }]=i(\Delta ^{\mu \lambda }S^{\nu \rho }-
\Delta ^{\mu \rho }S^{\nu \lambda }+\Delta ^{\nu \rho }S^{\mu \lambda }-
\Delta ^{\nu
\lambda }S^{\mu \rho })
\label{hurrato}
\end{equation}
with:
\begin{equation}
\Delta ^{\mu \nu }=\eta ^{\mu \nu }-\frac{a^{\mu }P^{\nu }+a^{\nu
}P^{\mu }}{P^0+m},
\end{equation}
that agrees with (\ref{dhat}) on the mass shell, and where
\begin{equation}
a^{\mu }=\eta ^{\mu 0}.
\end{equation}

We can enforce the Pryce constraints by using new coordinate and spin variables
$\hat{X } ^i$ and $\hat{S}^{ij}$, defined through the relations:
\begin{equation}
X^i-\hat{X
}^i=\frac{1}{P^0}(S^{0i}-\hat{S}^{0i}),
\label{XfraX}
\end{equation}
\begin{equation}
S^{ij}-\hat{S}^{ij}=-(X ^i-\hat{X}^i)P^j+(X ^j-\hat{X }^j)P^i.
\end{equation}
These relations ensure that the Lorentz generators $M^{\mu \nu }$ and therefore
also their commutation relations are unchanged. The Pryce constraints are:
\begin{equation}
\hat{S}^{0i}=\frac{1}{P^0}P^j\hat{S}^{ji}.
\end{equation}
Obviously:
\begin{equation}
P^i(X ^i-\hat{X }^i)=0,
\end{equation}
so:
\begin{eqnarray}
P^i(S^{ij}-\hat{S}^{ij})=\vec{P}^2(X ^j-\hat{X}^j)=
\frac{\vec{P}^2}{P^0}(S^{0j}-\hat{S}^{0j})=(P^0+m)S^{0j}-P^0\hat{S}^{0j}
\nonumber \\
\end{eqnarray}
i.e.:
\begin{equation}
(P^0+m)S^{0j}=m\hat{S}^{0j}
\label{transet}
\end{equation}
and thus:
\begin{equation}
X^i-\hat{X}^i=-\frac{1}{P^0+m}
\hat{S}^{0i},
\end{equation}
\begin{equation}
S^{ij}-\hat{S}^{ij}=\frac{1}{P^0+m}(P^j\hat{S}^{0i}
-P^i\hat{S}^{0j}).
\label{transto}
\end{equation}

Commutation relations are from (\ref{transet})-(\ref{transto}) combined with
(\ref{hurraet}) and (\ref{hurrato}) (with $\hat{X }^0$
commuting with everything):
\begin{equation}
[\hat{X }^{\mu },\hat{X
}^{\nu }]=\frac{i}{m^2}\hat{S}^{\mu
\nu }-\frac{i}{m^2P^0}(P^{\mu }\hat{S}^{0\nu }-P^{\nu }\hat{S}^{0\mu }),
\label{xwithx}
\end{equation}
\begin{eqnarray}
[\hat{X }^{\mu },\hat{S}^{\nu \lambda}]
=-\frac{iP^{\mu }}{m^2P^0}(P^{\nu
}\hat{S}^{0\lambda}-P^{\lambda }\hat{S}^{0\nu})+\frac{i}{m^2}(P^{\nu }
\hat{S}^{\mu
\lambda }-P^{\lambda }\hat{S}^{\mu \nu }).
\label{xwiths}
\end{eqnarray}

\end{document}